\documentclass[1p]{elsarticle}

\biboptions{sort&compress}

\usepackage[english]{babel}

\usepackage{amsfonts}
\usepackage{amsmath}
\usepackage{amssymb}
\usepackage{bbold}
\usepackage{bm}
\usepackage{braket}
\usepackage{hhline}
\usepackage{color,graphicx}
\usepackage{hyperref}
\usepackage{url}
\usepackage{ifpdf}
\usepackage[utf8]{inputenc}
\usepackage{slashed}

\usepackage{tikz-feynman}

\definecolor{blue}{rgb}{0.1,0.1,0.8}
\definecolor{nicered}{rgb}{0.7,0.1,0.1}
\definecolor{nicegreen}{rgb}{0.1,0.5,0.1}
\hypersetup{colorlinks,citecolor= blue,linkcolor= blue}

\newcommand{\beq}{\begin{equation}}
\newcommand{\eeq}{\end{equation}}
\newcommand{\bea}{\begin{eqnarray}}
\newcommand{\eea}{\end{eqnarray}}

\definecolor{Red}{rgb}{1.,0.,0.}

\newcommand{\lagr}{\mathcal{L}}

\newcommand{\lp}{\bm{p}_{\ell^+}}
\newcommand{\lm}{\bm{p}_{\ell^-}}
\renewcommand{\b}{\bm{p}_b}
\newcommand{\bbar}{\bm{p}_{\bar{b}}}
\newcommand{\h}{\bm{p}_h}

\newcommand{\kptld}{\tilde{ \kappa}}

\usepackage{ulem}

\usepackage{lineno,hyperref}
\modulolinenumbers[5]

\journal{Nuclear Physics B}

\begin{document}

\begin{frontmatter}

\title{Optimized probes of $CP$-odd effects in the $ t \bar t h$ process at hadron colliders}

\author[1]{Bla\v z Bortolato}
\ead{blaz.bortolato@ijs.si}
\author[1,2]{Jernej F.\ Kamenik}
\ead{jernej.kamenik@ijs.si}
\author[1,2]{Nejc Ko\v snik}
\ead{nejc.kosnik@ijs.si}
\author[1]{Aleks Smolkovi\v c \corref{cor1}}
\ead{aleks.smolkovic@ijs.si}

\address[1]{J. Stefan Institute, Jamova 39, P. O. Box 3000, 1001, Ljubljana, Slovenia}
\address[2]{Department of Physics, University of Ljubljana, Jadranska 19, 1000 Ljubljana, Slovenia}
\cortext[cor1]{Corresponding author}

\begin{abstract}
  We use machine learning~(ML) and non-ML techniques to study optimized $CP$-odd observables, directly and maximally sensitive to the $CP$-odd $i \kptld \bar t \gamma^5 t h$ interaction at the LHC and prospective future hadron colliders using the final state with a Higgs boson and a top quark pair, $pp\to t\bar t h$, followed by semileptonic $t$ decays. We perform phase-space optimization of manifestly $CP$-odd observables ($\bm \omega$), sensitive to the sign of $\kptld$, and constructed from experimentally accessible final state momenta. We identify a simple optimized linear combination $\bm \alpha\cdot \bm\omega$ that gives similar sensitivity as the studied fully fledged ML models. Using $\bm\alpha\cdot \bm\omega$ we project the expected sensitivities to $\kptld$ at HL-LHC, HE-LHC, and FCC-hh.
\end{abstract}

% arXiv:2006.13110

\end{frontmatter}
This article is registered under preprint number: /hep-ph/2006.13110

%%%%%%%%%%%%%%%%%%%%%%%%
%
\section{Introduction}
%
%%%%%%%%%%%%%%%%%%%%%%%%
The interaction between the heaviest particles of the Standard Model~(SM), the top quark $t$ and the Higgs boson $h$, is well known in the SM. The measured top quark mass $m_t$ and the electroweak condensate value $v$ precisely determine the on-shell scalar coupling $-y_t \bar t t h$ to be $y_t = \sqrt{2} m_t/v$, while the $P$- and $CP$-odd interaction $\bar t \gamma^5 t h$ is absent. Beyond the SM, effective operators of dimension-6 can break this correlation  and result in more general (pseudo)scalar $\bar t t h$ couplings $\kappa\ (\kptld)$~\cite{AguilarSaavedra:2009mx},
\begin{equation}
\label{eq:YukawaTopHiggs}
\begin{split}   
\lagr_{ht} &= -\frac{y_t}{\sqrt{2}} \bar t(\kappa  + i \kptld \gamma_5) t h\,,
\end{split}
\end{equation}
which reduce to the SM case at $\kappa = 1$, $\kptld = 0$. At the LHC it is possible to probe these couplings directly with two of the particles in Eq.~\eqref{eq:YukawaTopHiggs} on-shell\footnote{Since $m_h < 2m_t$ one cannot probe these couplings with all the three particles on-shell.} in top-Higgs associated production processes $p p \to t h j$ and $pp \to t \bar t h$~\cite{Ellis:2013yxa,Khachatryan:2016vau,Bhattacharyya:2012tj,hep-ph/9501339,  1407.5089, Boudjema:2015nda, 1507.07926, 1603.03632, Gritsan:2016hjl, Li:2017dyz, AmorDosSantos:2017ayi,1804.05874,Kobakhidze:2014gqa,1410.2701,1504.00611,1807.00281,Kraus:2019myc,hep-ph/9312210, Barger:2019ccj,Patrick:2019nhv,Ferroglia:2019qjy}\footnote{The loop induced partonic process $gg \to h \to t\bar t$ depends on $\kappa^2$, $\kappa$, and $\kptld^2$ already on the production side as it is dominated by the top quark loop~\cite{Brod:2013cka}.}. The corresponding total cross sections scale as $\kappa^2$, $\kptld^2$ ($thj$ also as $\kappa$), and are thus poorly sensitive to small nonzero $\kptld$. Linear sensitivity to $\kptld$ on the other hand can be achieved by measuring $P$- and $CP$-odd observables.

In a recent paper we have proposed {experimentally feasible and manifestly} $CP$-odd probes of $\kptld$ in $thj$ and $\bar t t h$ final states at the LHC and prospective future hadron colliders~\cite{Faroughy:2019ird}. The overwhelming irreducible backgrounds make the $thj$ channel impractical. For the $t \bar t h$ case we have identified 13 different $CP$-odd observables that can be constructed out of $5$ measurable final state momenta and an additional triple-product asymmetry~\cite{Durieux:2015zwa}. Namely, assuming $pp \to t \bar t h$ production with semileptonically decaying tops, we combined the final state lepton momenta $\lp, \lm$, two $b$-jet momenta $\b, \bbar$ from decaying top quarks (although without discriminating their charges\footnote{{Efficient $b$-jet charge discrimination could allow to construct further $CP$-odd observables, see e.g.~\cite{Bernreuther:1993hq, Boudjema:2015nda}.}}) and the Higgs momentum $\h$ in different ways to construct $C$-even, $P$-odd laboratory frame observables $\omega_i$~\cite{Faroughy:2019ird}. Note that the Higgs momentum $\h$ can be reconstructed in any feasible final state in the approach we propose. We have singled out the observable with the largest individual sensitivity to $\tilde \kappa$, namely
\begin{equation}
\label{eq:omega6}
\omega_6 = \frac{\left[(\lm \times \lp) \cdot (\b+\bbar)\right] \left[(\lm - \lp)\cdot (\b+\bbar)\right]}{|\lm \times \lp||(\b+\bbar)||\lm - \lp||\b+\bbar|}.
\end{equation}
In~\ref{app:omegas} we derive the set of $CP$-odd observables in a systematic way, completing the set introduced in Ref.~\cite{Faroughy:2019ird}, so that it now contains 22 $\omega$'s. Importantly, one among the new observables

\begin{equation}
\label{eq:omega14}
\omega_{14}= \frac{\left[(\lm \times \lp) \cdot (\b - \bbar)\right] \left[(\b - \bbar)\cdot (\lm-\lp)\right]}{|\lm \times \lp||\b - \bbar| |\b - \bbar||\lm-\lp|},
\end{equation}
gives similar performance to $\omega_6$.

 Due to the high dimensionality and complexity of the phase-space in this process with top quarks decaying semileptonically, in Ref.~\cite{Faroughy:2019ird} we have not ventured further in the search for an optimal CP probe of the $t\bar t h$ interaction. The aim of the present paper is to finally tackle this problem and use the complete kinematical information accessible experimentally to construct an optimal $CP$-odd observable. To this end we rely on neural networks~(NN) trained on Monte-Carlo generated samples to efficiently parametrize the weight function of events across the multi-dimensional phase-space in order to maximize the statistical sensitivity to $\tilde \kappa$. We show how the required $P$- and $CP$-symmetry properties of the NN-based observables can be imposed \textit{a priori}. Finally, we compare in terms of optimality, a general $CP$-odd NN function of the phase-space to a linear combination of manifestly $CP$-odd variables. {We note that existing general purpose ML inference tools are already able to optimize sensitivity to a given parameter (see e.g. Ref.~\cite{Brehmer:2019xox}). The purpose of this work is to do so in an economic way that manifestly respects the symmetries of the problem.}

The outline of this paper is as follows. In Sec.~\ref{sec:optimization} we perform the phase-space optimization of $\omega_6$ and $\omega_{14}$, analogous to the study of $th$ production in Ref.~\cite{Faroughy:2019ird}, but now applied to a multi-dimensional phase-space of semileptonic $t\bar t h$ process parametrized through a NN. Next, as a generalization to other available $\omega$'s we consider a manifestly $CP$-odd ($C$-even and $P$-odd) observable completely parameterized by a NN.
We also consider a first order approximation of this observable. In this limit, the significance optimization can be performed without the need for advanced machine learning techniques. At the same time we show that it is just slightly suboptimal compared to the fully fledged NN.
We use this optimized observable in Sec.~\ref{sec:bounds} to produce limits in the $\kappa-\kptld$ plane at HL-LHC \cite{HL-LHC-ApollinariG.:2017ojx,HL-LHC-CMS:2013xfa, HL-LHC-ATLAS:2013hta}, HE-LHC \cite{HE-LHC-Zimmermann:2018xtd, HE-LHC-Abada:2019ono}, and FCC-hh \cite{Mangano:2016jyj, Contino:2016spe,Benedikt:2018csr}. We conclude in Sec.~\ref{sec:conclusions}.

\section{Neural network approach to the optimal $CP$-odd observable in $t\bar t h$}
\label{sec:optimization}
We implement the training and evaluation of neural networks using the $\texttt{TensorFlow}$ framework~\cite{tensorflow2015-whitepaper}. In all cases, we use a sample of $10^7$ $p p \to h t(\to b \ell^+ \nu) \bar t(\to \bar b \ell^- \nu)$ events generated at LO using $\texttt{Madgraph5}$~\cite{Alwall:2014hca} together with the Higgs
Characterisation UFO model~\cite{Degrande:2011ua, Artoisenet:2013puc} with $\kappa,\kptld=1$. We split the sample into separate training (7.5M) and test (2.5M) samples. After training at fixed $\kptld=1$ we also test the observables at other values of $\kptld$ in this section and at other values of $\kappa$ in the following section, both ranging from $-1$ to 1. In these tests 1M events have been used, note that also a fresh sample for $\kptld=1$ was generated.
Unless stated otherwise the results are shown for events in $pp$ collisions at 14 TeV. We randomly initialize the neural network weights using the default Glorot uniform initializer and use the Adam optimizer with a custom varying learning rate $l(e)=l(e-1)/(1+0.8^e)$ where $e$ is the current epoch and the initial learning rate is set to $0.1$. {We use $\texttt{relu}$ for the activation function.} We train all networks using the loss function
\begin{equation}
\label{eq:loss_function_general}
\text{loss}(\bm{\alpha}) = \left(\frac{\text{mean}(\mathcal{F}(\bm{X}; \bm{\alpha}))}{\text{std}({\mathcal{F}(\bm{X}; \bm{\alpha})})/\sqrt{N}} \right)^{-2},
\end{equation}
where the $\text{mean}()$ and the standard deviation $\text{std}()$
are to be calculated over all events in the sample. The loss
corresponds to the inverse of the significance-squared of the
observable $\mathcal{F}(\bm{X}; \bm{\alpha})$ that should be minimized
in order to achieve optimal statistical sensitivity. Here $N$ is the
size of the sample, $\bm{\alpha}$ are the free neural network weights
and biases and $\bm{X}$ stands for the values of $CP$-even and/or
$CP$-odd phase-space variables in the given event. We avoid
over-fitting of the training sample by stopping the training when at
least 30 epochs have passed and one of the following two criteria is
satisfied: either the running average of 20 training losses saturates
to $0.5\%$ or the running average of 20 test losses increases for 5
epochs in a row. We keep a model history and in the end choose the
best model in terms of test loss. In practice we find that mostly the
first condition terminates the training loop, and the best model is
usually the model from the final epoch of training. In order to
determine the optimal NN architecture we perform a scan over a set of
possible NN configurations with up to 2 hidden layers and up to 9
nodes per NN layer.\footnote{
We have also considered an automated algorithm to determine the optimal
  NN architecture (i.e. Hyperopt~\cite{hyperopt},
  see also Ref.~\cite{Clavijo:2020mua} for one of its recent uses.).
  Here instead we present results of manual scans over a set of
  possible NN configurations in order to have better control over the
  NN parameters. We found the results of both approaches comparable.} {We choose this cutoff for representational purposes, however we have checked that our results do not change significantly when using larger networks, namely up to three hidden layers of 30 nodes each. }

\subsection{Phase-space optimization of a single $\omega$}
Here we study the optimization of the $\omega_6$ \eqref{eq:omega6} and $\omega_{14}$ \eqref{eq:omega14} variables  based on
phase-space averaging. We denote the $CP$-even phase-space variables with $\bm{x}$ and a single $CP$-odd one with $\omega_i$. Using this notation we can write the $t \bar t h$ production differential cross section with semileptonically decaying tops as
\begin{equation}
\label{eq:two-fold-x-sec}
\frac{d\sigma}{d\bm{x} d\omega_i} = A(\bm{x}, |\omega_i|)+\kptld \kappa B(\bm{x},\omega_i)\,.
\end{equation}
where $A$ is manifestly $CP$-even and $B$ a $CP$-odd function of
${\omega_i}$: {$B(\bm{x},{\omega_i}) = - B(\bm{x},-{\omega_i})$. The
$CP$- and $P$-odd part of the spectrum is due to interference of
scalar and pseudoscalar amplitudes, explaining the characteristic $\kappa \tilde \kappa$ dependence.} We do not follow the optimization
procedure based on separating $A$ and $B$ since this would require
cumbersome multidimensional binning~\cite{Atwood:1991ka}. We use a
vector of easily accessible $CP$-even Mandelstam variables $\bm{x}$:
\begin{equation}
\bm{x} = \begin{pmatrix} &(\lp + \lm) \cdot \h\\
&(\lp + \lm) \cdot (\b + \bbar)\\ & (\b + \bbar) \cdot \h\\ & \lp \cdot \lm\\ & \b \cdot \bbar
\end{pmatrix}.
\end{equation}
Our goal is to find the optimal $CP$-even weight function $f(\bm{x};\bm{\alpha})$,
which should be used to calculate the weighted average of ${\omega_i}$. The function $f$ takes $CP$-even quantities $\bm{x}$ as inputs, therefore we expect its dependence on $\kptld$ to be of the form
\begin{equation}
f(\bm{x};\bm\alpha) = C(\bm{x};\bm{\alpha}) + \kptld^2 D(\bm{x};\bm\alpha) + \mathcal{O}(\kptld^4)\,.
\end{equation}
Using \eqref{eq:two-fold-x-sec} we can now express the observable
\begin{equation}
\label{eq:phase-space-w6}
\begin{aligned}
\braket{f(\bm{x};\bm\alpha) {\omega_i}} &= \int \frac{d\sigma}{d\bm{x} d{\omega_i}} \,f(\bm{x};\bm\alpha) {\omega_i} \,d\bm{x} d{\omega_i} \\
 &= \kptld \kappa \int B(\bm{x},{\omega_i}) C(\bm{x};\bm\alpha) \omega\, d\bm{x} d{\omega_i} + \kptld^3 \int B(\bm{x}, {\omega_i}) D(\bm{x};\bm\alpha)\, d\bm{x} d{\omega_i} + \mathcal{O}(\kptld^5)\,,
\end{aligned}
\end{equation}
{with the usual definition of the average.}\footnote{{The phase space average of a function is defined as $\braket{\#} \equiv\int  \tfrac{d\sigma}{d\bm{x} d\omega} \#\, d\bm{x} d\omega$.}}
The presence of odd powers of $\kptld$ reflects the $CP$-oddness of the observable. The large dimensionality of the phase-space suggests the parameterisation of the function $f(\bm{x};\bm{\alpha})$ by means of an appropriate NN. In terms of the loss function \eqref{eq:loss_function_general} we have $\mathcal{F}(\bm{x},{\omega_i}; \bm{\alpha}) = f(\bm{x}; \bm{\alpha})~{\omega_i}$.

\begin{figure}[!h]
  \centering
  \begin{tabular}{c}
    	\includegraphics[scale=0.79]{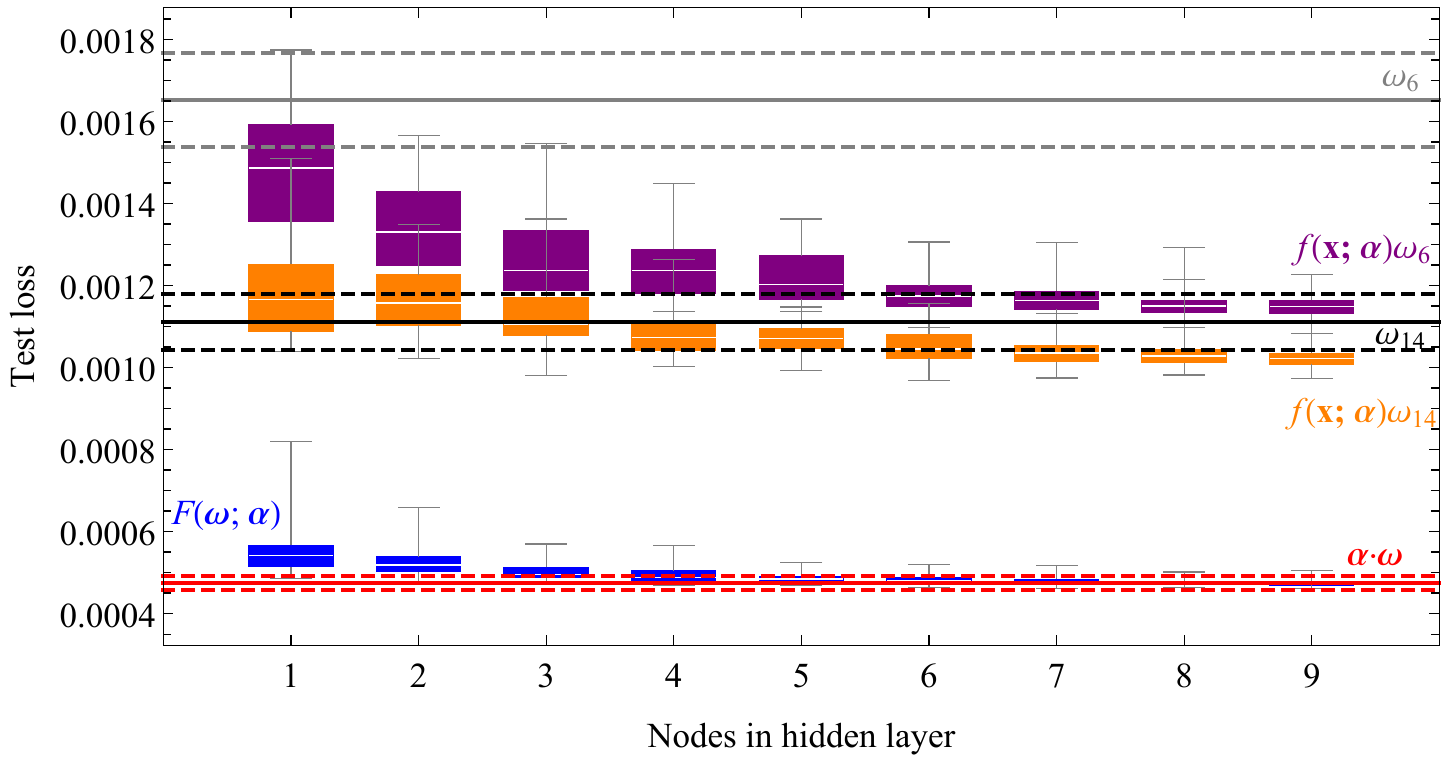}\\
	\includegraphics[scale=0.34]{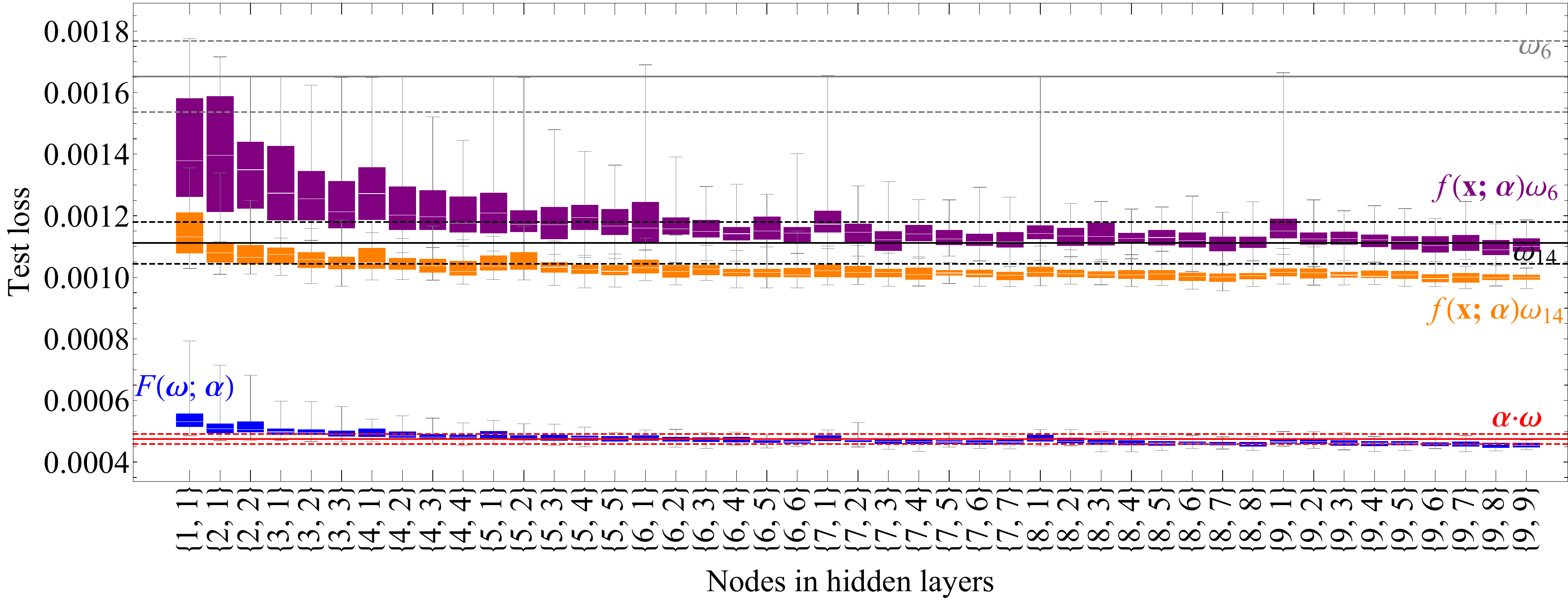}
  \end{tabular}
	\caption{A scan in terms of the test loss (sample size 2.5M, {$\kappa,\kptld=1$}) over neural network configurations with one (upper plot) or two (lower plot) hidden layers for the phase-space optimized $\omega_6$ {and $\omega_{14}$} \eqref{eq:phase-space-w6} shown in the purple {and orange} box plot and the generalized $F(\bm{\omega};\bm{\alpha})$ (Sec.~\ref{neural-net-CP-odd}) shown in the blue box plot. The spread in {all} cases corresponds to 50 different random weight initializations per configuration. For comparison the plain $\omega_6$ \eqref{eq:omega6} {and $\omega_{14}$ \eqref{eq:omega14} are} shown in gray {and black} with the dashed lines showing {their} $1\sigma$ statistical uncertainty. The first order approximation of $F(\bm{\omega}; \bm{\alpha})$, defined in Eq.~\eqref{eq:constAlpha}, is shown in red as described in Sec.~\ref{first-order-approx}.}
	\label{fig:hidden_comp}
\end{figure}

To understand the impact of using different possible neural network architectures, we have performed a manual scan over a set of neural network configurations. The input layer has 5 nodes (one per each $\bm{x}$ component) and the output layer has one node resulting in a scalar $f(\bm{x};\bm{\alpha})$. We study networks with a single hidden layer of 1-9 nodes and double hidden layer networks with 1-9 nodes each, constraining the number of nodes on the second hidden layer to be smaller than or at most equal to the number of nodes on the first hidden layer. The results of the converged test losses of 50 different random weight initializations per configuration are shown on Fig.~\ref{fig:hidden_comp} in the purple box plot {for the case of $\omega_6$ and the orange box plot for the case of $\omega_{14}$}. The plain $\omega_6$-based observable is shown in gray, with the dashed lines denoting its $1\sigma$ statistical uncertainty{, while the same holds true for plain $\omega_{14}$ in black}. We find that the phase-space optimization of $\omega_6$ gives a noticeable improvement over plain $\omega_6$ when using a large enough network. {Interestingly, $\omega_{14}$ seems to be close to optimal on its own, as the phase space optimization does not introduce noticeable improvement. Moreover, the optimized $\omega_6$ gives a similar performance to $\omega_{14}$, hinting that we have reached maximal performance achievable with a single $\omega$}.

To test how well the resulting {networks generalize} to other values of $\tilde{\kappa}$ we use the 50 converged $\{9,9\}$ models and calculate the dependence of the resulting {observable significances} with respect to $\tilde{\kappa}$ {on the afforementioned sets of 1M events per $\tilde{\kappa}$}. This is shown on Fig.~\ref{fig:sig_kptld} where a consistent improvement over simple $\langle\omega_6\rangle$ can be seen at all considered $\tilde{\kappa}${, while again a marginal improvement is confirmed for $\omega_{14}$, with the optimized $\omega_6$ hovering around $\omega_{14}$}.

{As the results of optimizing single $\omega_6$ and $\omega_{14}$ point to a maximal performance possible using a single $\omega$ and the chosen set of phase space variables, we now turn to the rest of the $\omega$'s. In the next subsection we consider a more general case where the CP-odd observable itself is parameterized with a neural network.}

\subsection{Neural network as a $CP$-odd observable}
\label{neural-net-CP-odd}
Here we consider a case where the output of the neural network is a $CP$-odd quantity that defines our observable. We build a network with $14$ inputs, one per each $\omega_i$, and one output $F(\bm\omega; \bm{\alpha})$, which is correctly anti-symmetrized so that $F(\bm\omega; \bm{\alpha}) = - F(-\bm\omega; \bm{\alpha})$. The loss function defined in Eq.~\eqref{eq:loss_function_general} is now $\mathcal{F}(\bm{X}; \bm{\alpha}) = F(\bm\omega; \bm{\alpha})$. {Note that since we include the complete irreducible set $\bm \omega$ in this non-linear construction, it effectively also covers the case of a simple phase-space optimization of any (linear combination of ) $\omega_i$, since all relevant CP-even phase-space variables can be recovered by taking suitable ratios of $\omega_i/\omega_{i'}$.}

We again carry out the study of the dependence of the network size with respect to the test sample loss, including non-negligible uncertainties associated with random weight initializations. We scan the neural network architecture parameter space in the same way as in the previous case, starting with a single hidden layer of 1-9 nodes, then adding an additional hidden layer with the number of nodes smaller than or equal to the number of nodes on the first hidden layer. For each configuration we run 50 trainings with different random weight initializations. The results are shown in Fig.~\ref{fig:hidden_comp} in the blue box plot. We find a considerable improvement over the phase-space optimizations of {$\omega_6$ or $\omega_{14}$. The improvement is consistent in the entire range of $\kptld \in [0.1, 1.0]$ and is most striking at large $\kptld$.}

\begin{figure}[!h]
	\centering
	\includegraphics[scale=0.8]{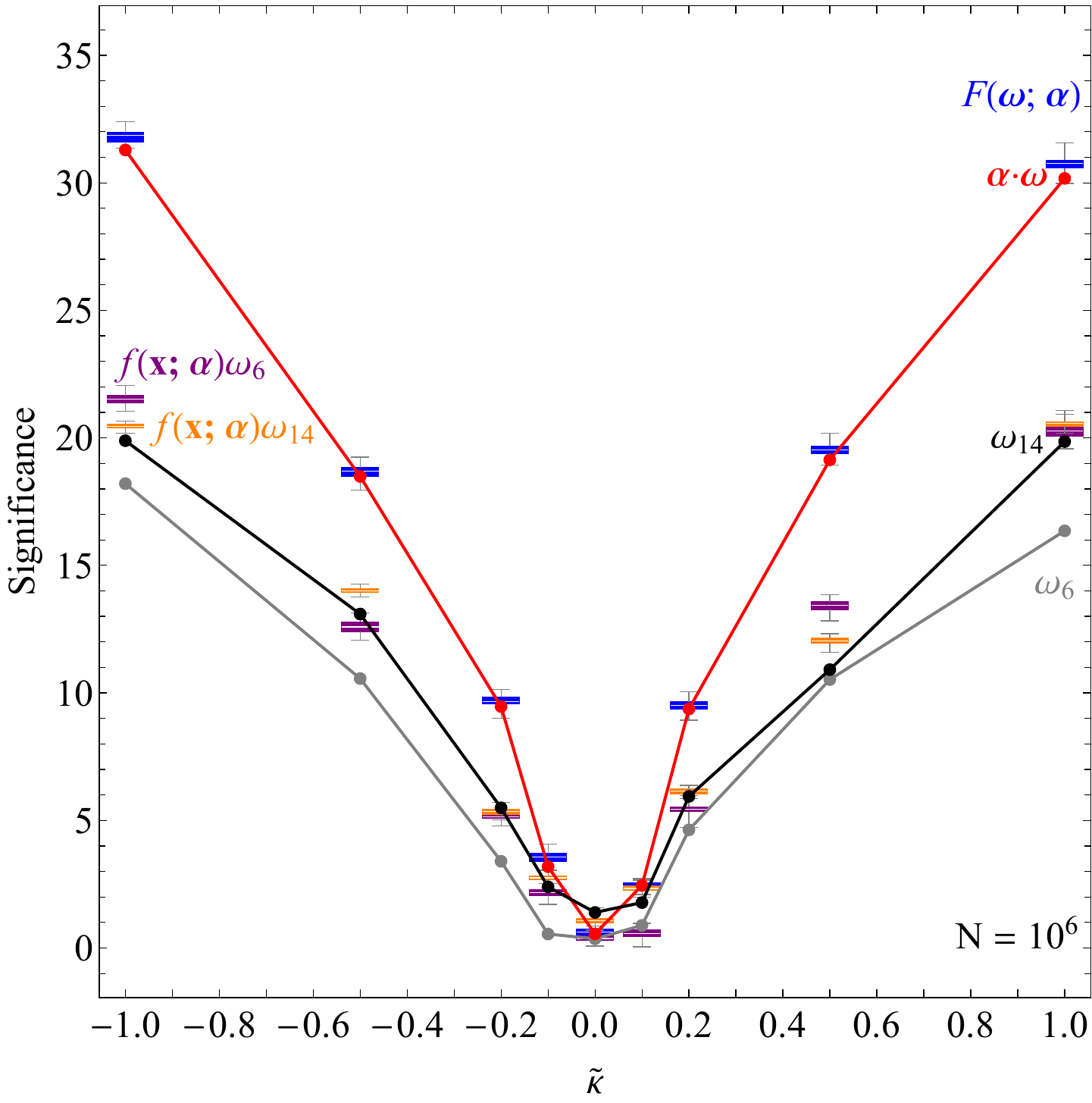}
	\caption{
		Comparison of the significances {(defined as the mean value divided by the standard deviation)} of all the observables considered in this work with respect to $\kptld$ {and at fixed $\kappa=1$}. The results correspond to 1M events per $\kptld$ at 14 TeV. Plain $\omega_6$ \eqref{eq:omega6} in gray, $\omega_{14}$ \eqref{eq:omega14} in black, phase-space optimized $\omega_6$ {and $\omega_{14}$} \eqref{eq:phase-space-w6} in purple {and orange}, anti-symmetrized neural network $F(\bm \omega; \bm \alpha)$ (Sec.~\ref{neural-net-CP-odd}) in blue and the first order approximation of the latter $\bm \alpha \cdot \bm \omega$ in red (Sec.~\ref{first-order-approx}). See text for details on each observable.}
	
	\label{fig:sig_kptld}
\end{figure}

Again we check the generalizing power of the resulting observables to other $\tilde{\kappa}$ by fixing the model configuration to $\{9,9\}$ and calculating the significance of the resulting observables with respect to $\kptld$. The results are shown on Fig.~\ref{fig:sig_kptld}. We find a consistent improvement over the previous case across all considered $\kptld$. A noticeable improvement in the significance can be seen.
{In order to better understand the physics underlying the optimization}, we next consider this model in the leading order approximation in $\bm\omega$.

\subsection{First order approximation of $F($\boldmath$\omega$\unboldmath$;$\boldmath$\alpha$\unboldmath$)$}
\label{first-order-approx}
 \begin{figure}[!h]
	\centering
	\includegraphics[scale=0.525]{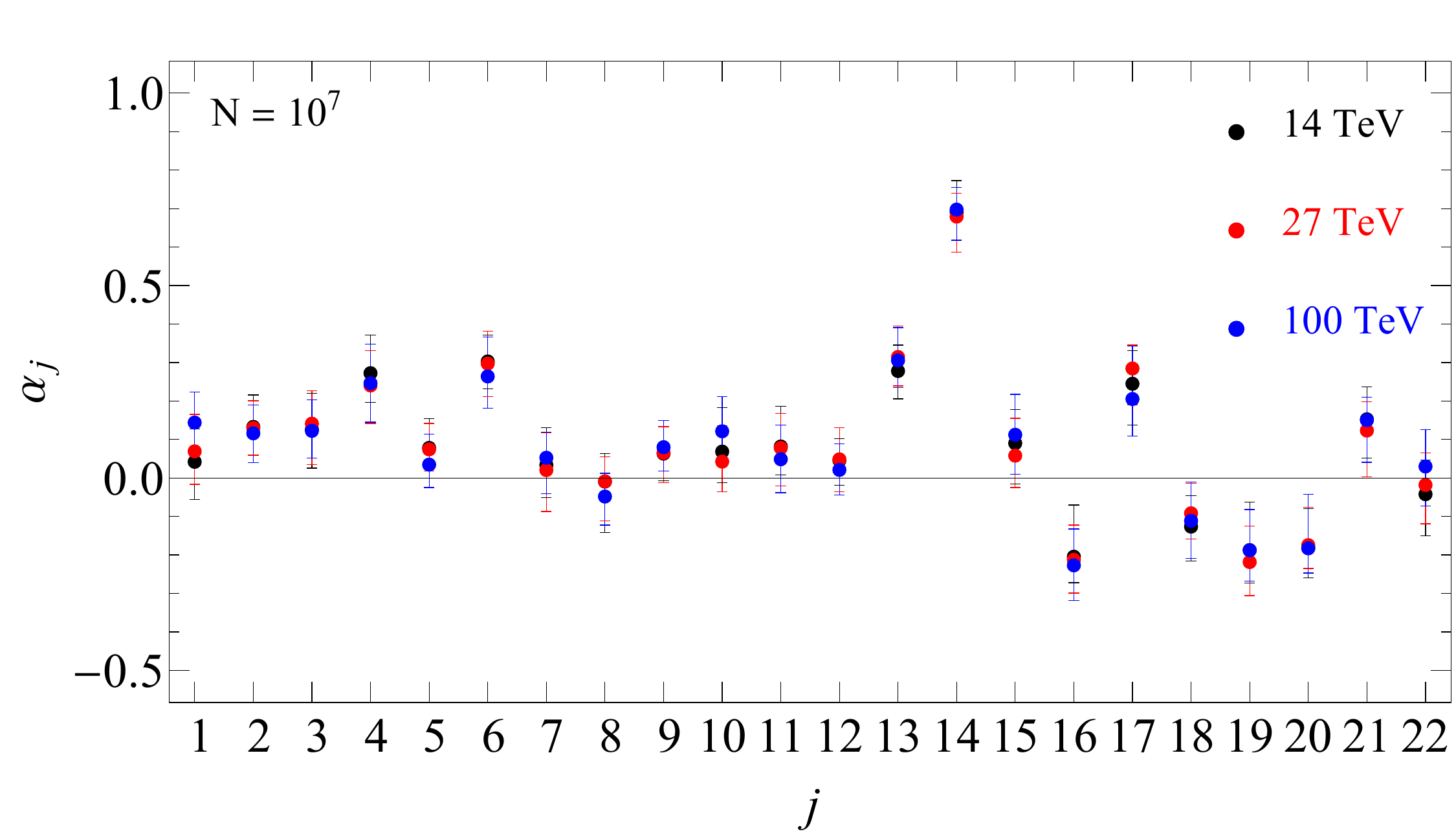}~~~~~ \\
	\includegraphics[scale=0.5]{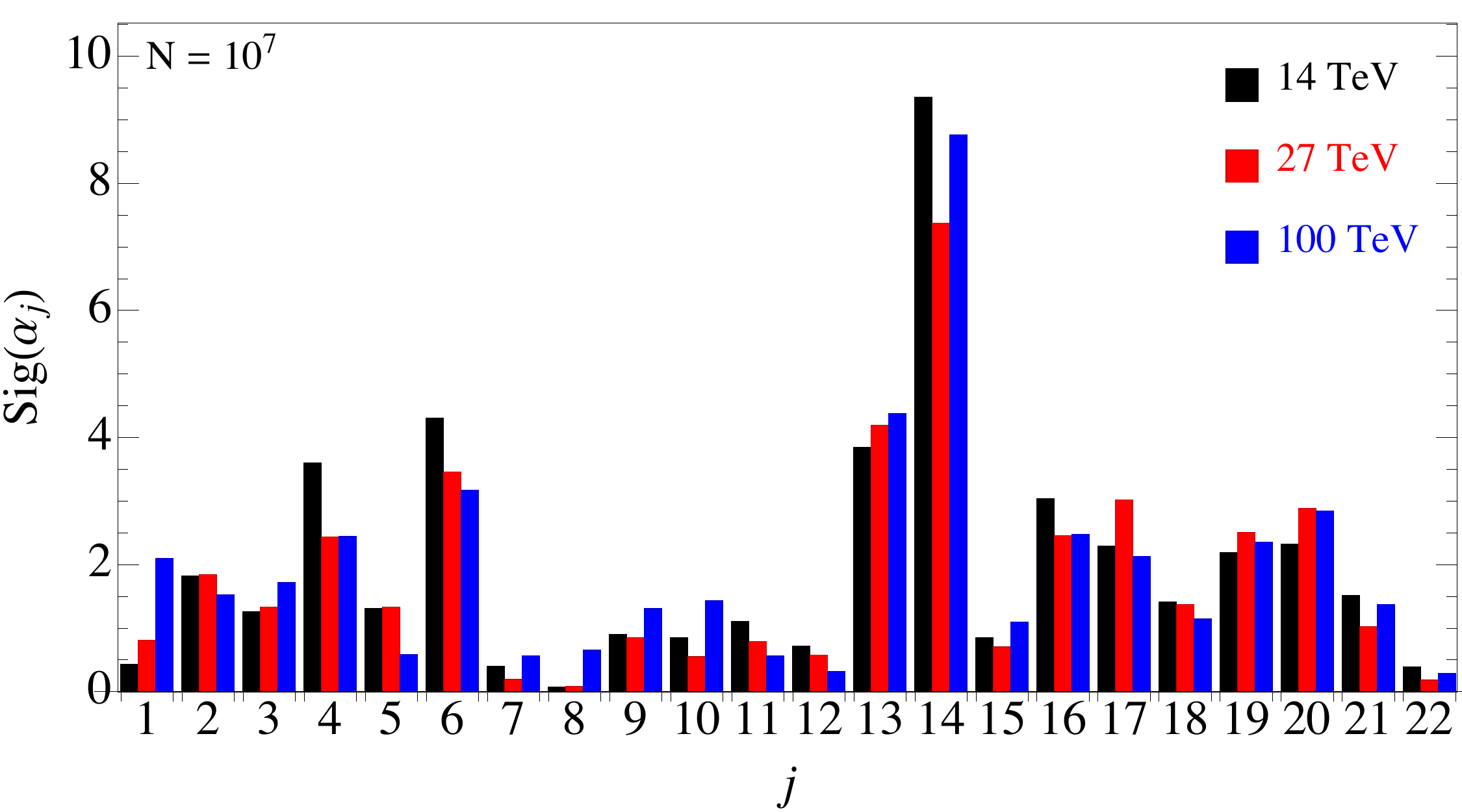}
	
	\caption{Optimal weights of the linear observable defined in Eq.~\eqref{eq:constAlpha}. The upper plot shows uncertainties of $\alpha_j$, estimated using the expected statistical errors of the observable significances, see text for details. The lower plot shows the significances of $\alpha_j$ (defined as their central value divided by their estimated uncertainty). }
	\label{fig:opti_alphas_energ}
\end{figure}
To address the arbitrariness of the neural network architecture choice and 
{to better understand the underlying physics},
in this Section we consider the first order $\bm\omega$ approximation of the form of $F(\bm\omega; \bm{\alpha}) = \sum_j \alpha_j \omega_j + \mathcal{O}(\omega^3)$ for $j  \in \{1,\ldots,22 \}$. The approximation is justifiable in terms of a Taylor expansion, as most of the events have $|\omega_j| \ll 1$\,.\footnote{Note that $|\omega_i| < 1$ by definition, whereas in some cases (also for $\omega_6$) the upper bound is $1/2$. See~\ref{app:omegas}.} The observable is then simply
\begin{equation}
\label{eq:constAlpha}
\bm{\alpha}\cdot \bm{\omega} = \braket{\sum_j\alpha_j \omega_j}\,,
\end{equation}
with the subsidiary condition $|\bm{\alpha}| = 1$.
We can optimize over $\alpha_j$ by maximizing the significance
\begin{equation}
  \label{eq:SigStationary}
\frac{\partial }{\partial \alpha_j}
\frac{\bm{\alpha}\cdot \bm{\omega}}{
	\text{std}(\bm{\alpha}\cdot \bm{\omega})}
 = 0\,.
\end{equation}
Doing so we obtain a system of 22 quadratic equations \footnote{Notice that the problem is equivalent to a single neuron NN with 22 inputs and one output without the activation function or the bias term.}
\begin{equation}
\alpha^T M^{(j)} \alpha = 0\,,
\end{equation}
 \noindent
where $ \alpha = [\alpha_1,\ldots, \alpha_{22}]^T $ and $22 \times 22$ matrices $M^{(j)}$ are given by
\begin{equation}
M^{(j)}_{ik} = \langle \omega_i  \omega_j \rangle \langle \omega_k \rangle
 - \langle \omega_i  \omega_k \rangle \langle \omega_j \rangle. 
\end{equation}
We use this approach to extract the optimal weights $\alpha_j$ from $10^7$ events generated with $\kptld = {\kappa = 1}$ at 14, 27, and 100 TeV. We estimate the uncertainty associated with the optimal weights in the following way. First we estimate the statistical spread of the significance obtained with optimal $\bm\alpha$. Next we allow a single $\alpha_j$ to float in the intervals $[\alpha_j-\sigma_j, \alpha_j+\sigma_j]$, where $\sigma_j$ is chosen such that the decrease of the significance due to the change in $\alpha_j$ corresponds to the statistical spread of the significance. We perform an efficient scan around the optimal vector $\bm\alpha$ in its {22}-dimensional neighborhood using spherical coordinates to trivially fulfill the normalization constraint $\sum_j \alpha_j^2 = 1$. We approximate the significance with a quadratic function around the extremum to find independent, uncorrelated directions in the $\alpha$-space. With this procedure we determine how sharply the optimal $\alpha_j$ are defined. We estimate the statistical error of the significance using $10^7$ events. Clearly the uncertainties $\sigma_j$ are larger for smaller chosen sample size. {The results of this approach are shown in Fig.~\ref{fig:opti_alphas_energ}, where the upper (lower) panel shows the estimated error (significance) for each $\alpha_j$ at 14, 27, and 100 TeV.} {A comparison of the observable $\bm{\alpha}\cdot \bm{\omega}$ to other approaches in this work is shown in Fig.~\ref{fig:sig_kptld}. We reach a similar level of improvement compared to the full $F(\bm \omega; \bm \alpha)$ network with significantly fewer parameters.}

\begin{figure}[!h]
	\centering	\includegraphics[scale=0.3]{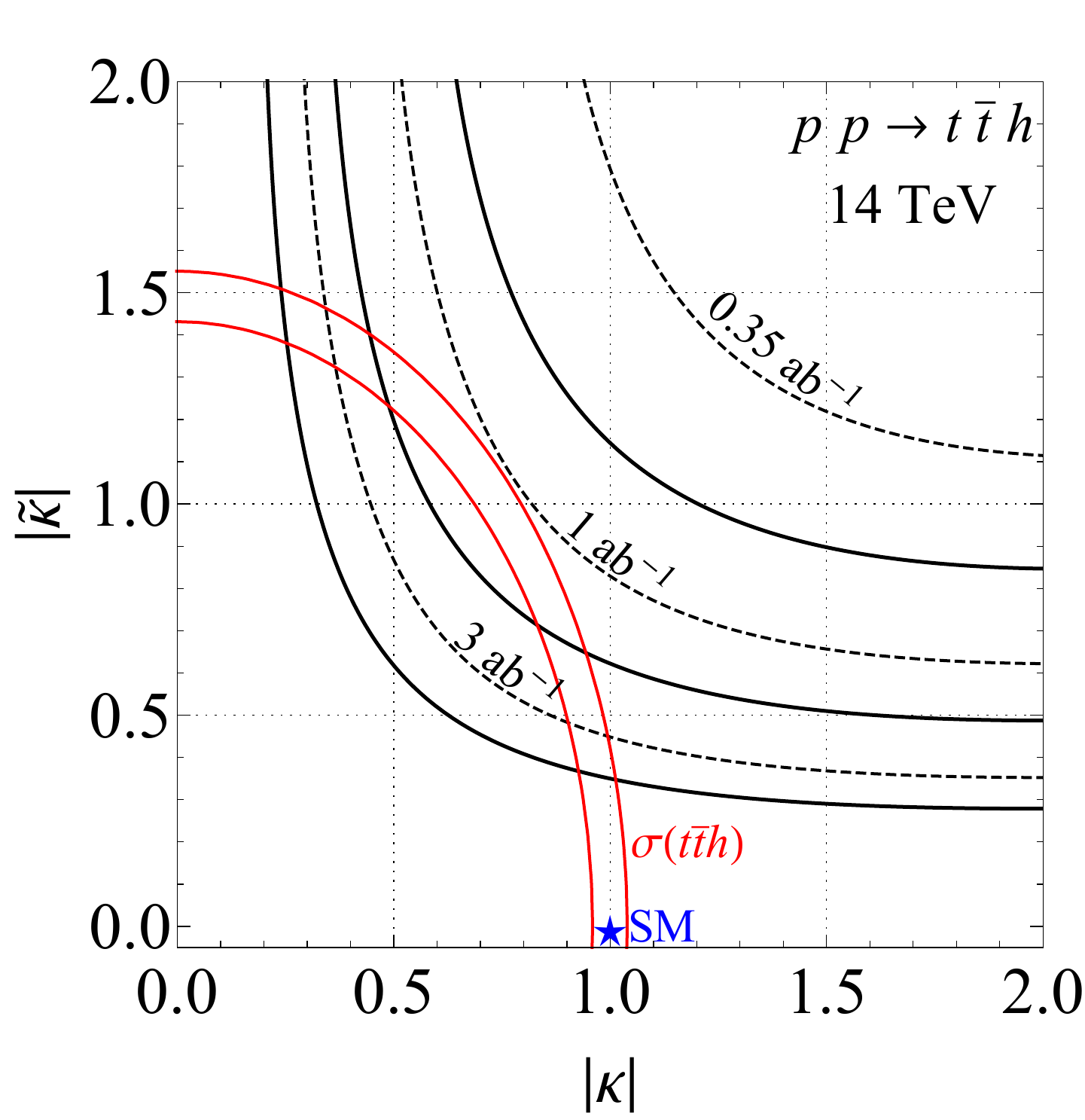}~\includegraphics[scale=0.3]{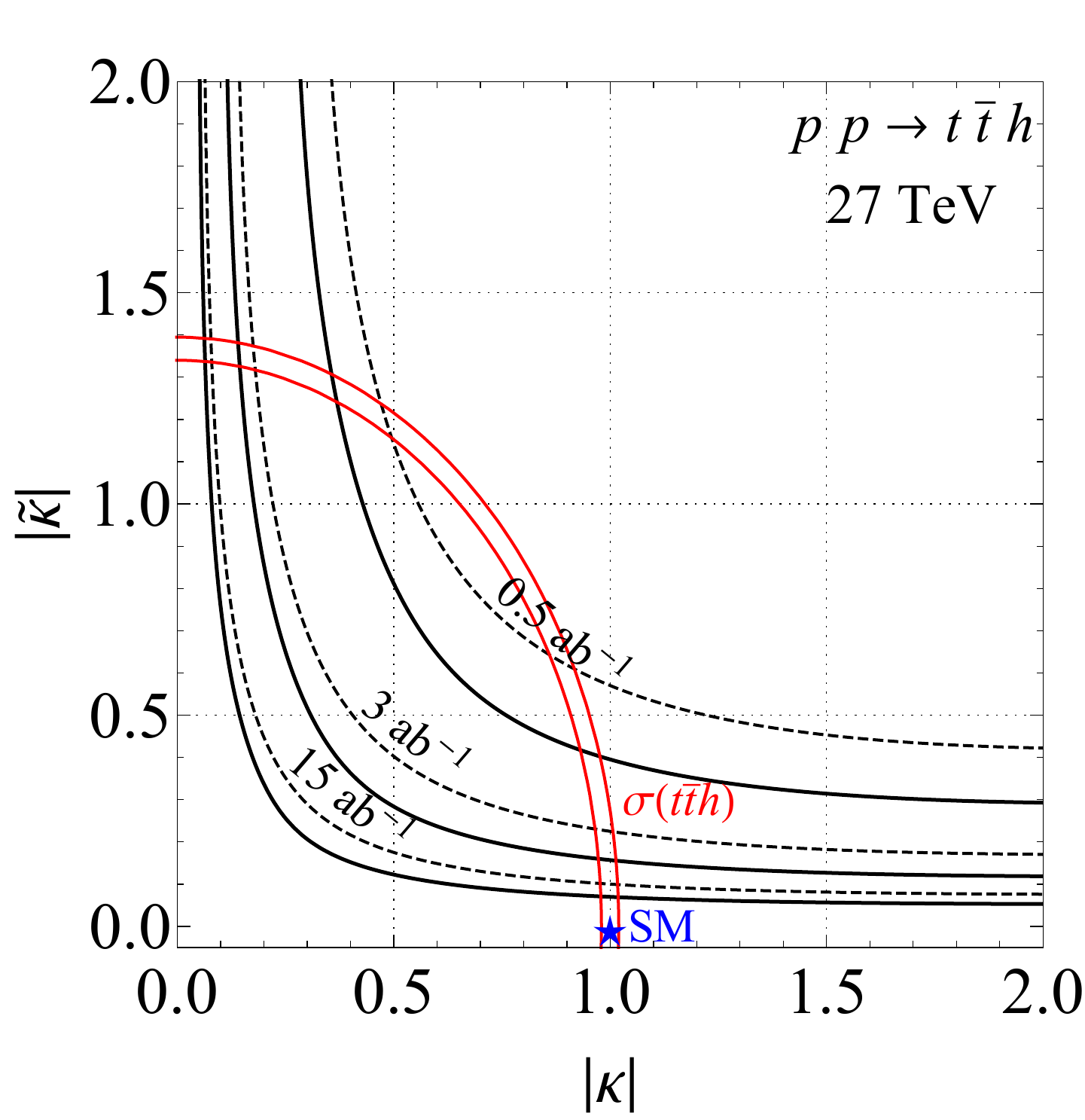}~
	\includegraphics[scale=0.3]{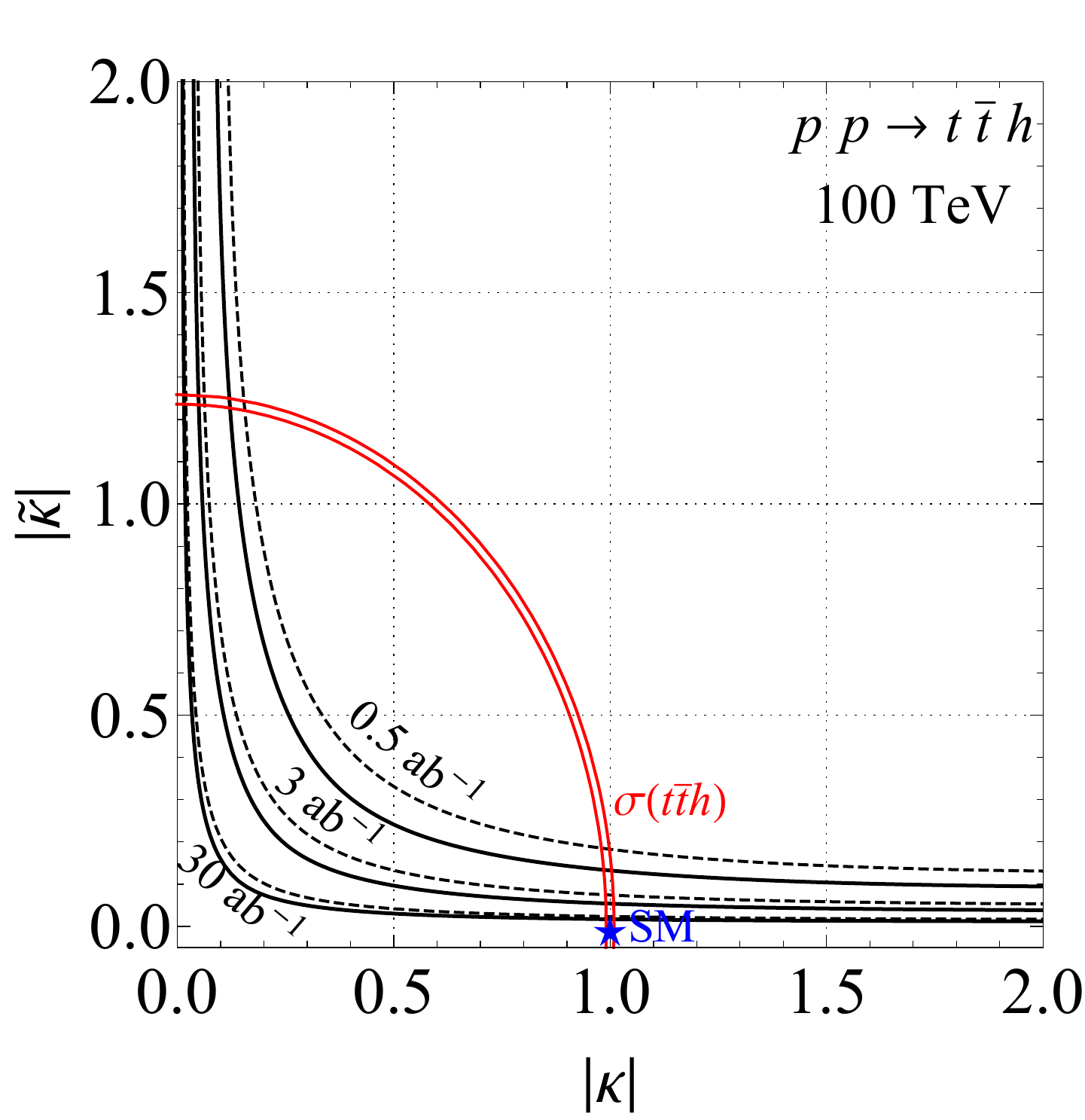}
	\caption{The $2\sigma$ exclusion zones in the $\kappa-\kptld$ plane by assuming a null result at HL-LHC, HE-LHC and FCC-hh for different luminosities. The optimized observable $\bm{\alpha}\cdot \bm{\omega}$ is shown in solid black, while the plain $\omega_6$ \eqref{eq:omega6} {(for direct comparison with \cite{Faroughy:2019ird})} results are shown using dashed lines. {The projected sensitivity of $t\bar t h$ production cross-section measurements is shown in red, see text for details}. At 14 TeV order 1 exclusion {of $\kptld$} can be achieved {using $\bm{\alpha}\cdot \bm{\omega}$} with $350\,\text{fb}^{-1}$ which corresponds to the final integrated luminosity of the LHC. 
}
	\label{fig:tth-lums}
\end{figure}

\begin{figure}[!h]
	\centering
	\includegraphics[scale=0.6]{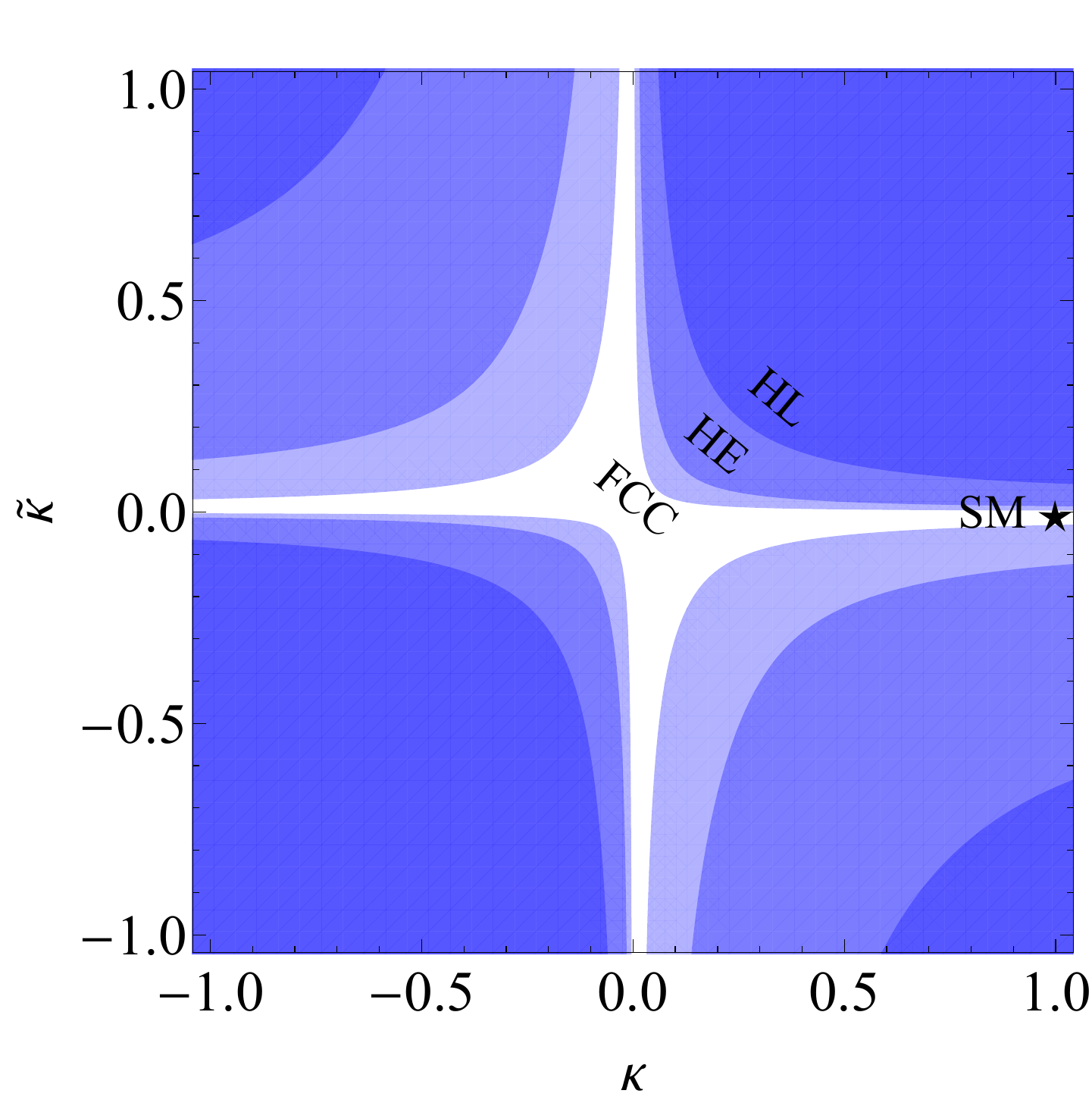}
	\caption{The $2\sigma$ exclusion regions at HL-LHC (3 ab$^{-1}$), HE-LHC (15 ab$^{-1}$) and FCC-hh (30 ab$^{-1}$) by assuming a measurement of a $2\sigma$ positive fluctuation in the optimal observable $\bm{\alpha}\cdot \bm{\omega}$ \eqref{eq:constAlpha}. }
	\label{fig:asym_bounds}
\end{figure}

\section{Bounds in the $(\kappa, \tilde{\kappa})$ plane}
\label{sec:bounds}
We produce the bounds in the $(\kappa, \tilde{\kappa})$ plane by {varying both $\kappa$ and $\kptld$ at generator level and} including showering and hadronization effects using $\texttt{Pythia8}$ and detector effects using $\texttt{Delphes}$ with the default ATLAS simulation card. As the $t\bar t h$ is followed by semileptonic top decays and $h \to b \bar b$ decay, our signal is defined as 4 $b$-jets and two oppositely charged leptons $\ell$.
{We  include} the main irreducible background $pp \to t \bar{t} b \bar{b}$ with both tops decaying semileptonically and use the same event selection requirements as in \cite{Faroughy:2019ird}:
{
\begin{itemize}
	\item $\geq 4$ jets of any flavor with $|\eta(j)|<5$ and $p_T (j) > 20$ GeV.
	\item $\geq 3$ of the above jets are $b$-tagged.
	\item 2 oppositely charged light leptons with $|\eta(\ell)|<2.5$ and $p_T(\ell)>10$~GeV.
\end{itemize}
 Furthermore, in order to identify the $b$-jets from $t$, $\bar t$ decays we count the number $N_b$ of tagged $b$-jets and perform the following selections: if $N_b\ge 4$, we compute the invariant masses $m_{bb}$ of all possible $b$-jet pairs and select the pair with invariant mass closest to the Higgs mass $m_h=125$ GeV. If the selected pair falls inside the Higgs mass window ($|m_{bb} - m_h| < 15$ GeV) we remove the pair from the list of $b$-jets and select from this list the highest $p_T$ $b$-jets as our candidate top quark decay $b$-jets. However, if $N_b=3$ we compute all possible invariant masses $m_{bj}$ where $j$ are non-$b$ jets in the event. We select as the $h\to b\bar b$ candidate the $bj$ pair that minimizes $|m_h-m_{bj}|$ and falls inside the Higgs mass window $m_h\pm 15$ GeV. The remaining two $b$-jets are taken as the candidate top quark decay $b$-jets.
}

We {update the} bounds for HL- and HE-LHC and produce bounds for FCC-hh for the first time by using the optimal observable \eqref{eq:constAlpha} {with the weights shown in the upper plot of Fig.~\ref{fig:opti_alphas_energ}}. The bounds coming from a null result up to the expected statistical uncertainty for different luminosities at different energies are shown on Fig.~\ref{fig:tth-lums}, {where we also show the expected sensitivity to $(\kappa, \kptld)$ from the $t\bar t h$ production cross-section measurements using the projected uncertainties of $\delta\kappa/\kappa \sim (0.04, 0.02, 0.009)$ at HL-LHC, HE-LHC~\cite{Cepeda:2019klc} and FCC-hh~\cite{Abada:2019lih}, respectively. {For direct comparison with \cite{Faroughy:2019ird} we also show the expected bounds from using a single $\omega_6$ and have checked that using $\omega_{14}$ does not change the single $\omega$ bounds significantly.}} 

A consistent improvement of sensitivity can be achieved by using the optimized combination of $\omega$'s with respect to a single $\omega_6$, {constraining the parameter space significantly in orthogonal directions compared to the cross-section measurements}. Interestingly the significance improvement is consistent between partonic events and after including shower, detector effects, {as well as the dominant background and realistic object reconstruction}, even though the optimization was performed at parton level only. This robustness is a welcome benefit of the method, since the computationally costly optimization procedure does not appear to be sensitive to modeling of the hadronic final states and detector effects. It is also reassuring that the optimization does not significantly rely on specific phase-space regions particularly affected by the background. {We expect the results to be also robust against higher order QCD corrections. In Ref.~\cite{Faroughy:2019ird} we have explicitly compared how the inclusion of NLO QCD effects changes the optimized observables in the case of $p p \to t h j$  and found a shift of the optimal weight function smaller than $10\%$ {and essentially negligible effect on the final sensitivity. In the present case, we have checked explicitly that weights optimized on parton level events can be applied to fully showered and reconstructed objects and still give close to optimal sensitivity. We thus deem our method robust with respect to theoretical and experimental systematics.} }

We show the sensitivity of the optimized observable to the sign of $\kptld$ (and $\kappa$) on Fig.~\ref{fig:asym_bounds} by assuming the measurement of a $2\sigma$ positive statistical fluctuation of the SM case, which in our estimate corresponds to the measurement of $\bm\alpha \cdot \bm\omega = (4.6 \pm 2.3) \times 10^{-4}$, $\bm\alpha \cdot \bm\omega = (0.9 \pm 0.45) \times 10^{-4}$ and $\bm\alpha \cdot \bm\omega = (0.2 \pm 0.1) \times 10^{-4}$ for HL-LHC (3 ab$^{-1}$), HE-LHC (15 ab$^{-1}$) and FCC-hh (30 ab$^{-1}$) respectively.

\section{Summary and conclusions}
\label{sec:conclusions}
{We revisited a set of manifestly $CP$-odd observables ($\omega_i$) built from experimentally accessible final state momenta in $p p \to t \bar t h$ production with semileptonicaly decaying tops.} In particular, we studied the prospect of their phase-space optimization, parameterizing the optimal weight functions with neural networks. First we considered the phase-space optimization of single $\omega_6$ {and $\omega_{14}$, showing that the performance of $\omega_6$ can be optimized, reaching the level of using a single $\omega_{14}$, which is already close to optimal.} Next we studied a general $CP$-odd observable, parameterized directly by an anti-symmetric neural network, which resulted in better performance. Lastly, we studied the first order approximation of this network as a linear combination of the $CP$-odd observables, producing a simpler and more robust observable, $\bm \alpha \cdot \bm \omega$. One benefit of using $\bm \alpha \cdot \bm \omega$ optimized at parton level is that it retains close to optimal sensitivity to the CP-odd coupling even after detector simulation, event selection and reconstruction, allowing to probe $\tilde{\kappa}$ directly at HL-LHC, HE-LHC and FCC-hh. We found that, at the end of Run 3, the LHC will exclude $\kappa \kptld \sim 1$ with $2\sigma$ confidence, while FCC-hh will be sensitive to $\kappa \kptld \sim 0.01$\,.
{Note that these observables represent highly complementary probes of the top Yukawa sector compared to $pp \to t\bar t h$ cross-section measurements. In particular in their optimized form they would allow to break the degeneracy in the $(\kappa,\tilde\kappa)$ plane and significantly reduce the allowed parameter space even at modest sensitivities accessible at the (HL)LHC.} 
Finally, our approach to parametrizing $CP$-odd observables over high-dimensional phase-spaces using manifestly $CP$-odd NNs could be applied to other high energy particle production and decay processes, as well as to other symmetries. We leave the exploration of these ideas for future work.

\section*{Acknowledgments}
The authors acknowledge the financial support from the Slovenian Research Agency (research core funding No. P1-0035 and J1-8137). This article is based upon work from COST Action CA16201 PARTICLEFACE supported by COST (European Cooperation in Science and Technology). A.~S. is supported by the Young Researchers Programme of the Slovenian Research Agency under the grant No. 50510, core funding grant P1-0035.

\appendix

\section{Construction of $\omega_1,\ldots,\omega_{22}$}
\label{app:omegas}
In this Section we derive the most general set of linearly independent pseudoscalars of mass-dimension 5 that are even under charge conjugation and under $b \leftrightarrow \bar b$. It is convenient to start from 5 independent 3-vectors with well-defined $C$: $V \in \{\h, \lm+\lp,\lm-\lp,\b+\bbar,\b-\bbar\}$. All of the quintuple products with desired properties can be expressed as $\omega \sim V_1 \times V_2\cdot V_3\, V_4 \cdot V_5$.\footnote{Notice that the possibility of a nested cross product $(((V_1\times V_2) \times V_3)\times V_4)\cdot V_5$ can also be reduced to this form.}
We can find systematically all $\omega$'s by considering $10$ distinct mixed products $V_1 \times V_2 \cdot V_3$. Multiplying these together with $15$ different $V_4 \cdot V_5$ we have 150 potential quintuple products. In the last step we symmetrize the obtained product with respect to $C$-conjugation and $b \leftrightarrow \bar b$ and we find that only a handful of $\omega$'s remain, which we list below.

In order to reduce the number of linearly independent $\omega$'s we have employed an Euclidean tensor identity
\begin{equation}
  \label{eq:EpsIdentity}
  \delta_{ab} \epsilon_{cde} - \delta_{ac} \epsilon_{deb} + \delta_{ad} \epsilon_{ebc} - \delta_{ae} \epsilon_{bcd} = 0\,,
\end{equation}
or, equivalently, that the following linear combination of four arbitrary vectors vanishes:
\begin{equation}
  \label{eq:VectorID}
  \bm{a}\,(\bm{b}\times \bm{c} \cdot \bm{d})-  \bm{b}\,(\bm{c}\times \bm{d} \cdot \bm{a}) +
    \bm{c}\,(\bm{d}\times \bm{a} \cdot \bm{b}) -   \bm{d}\,(\bm{a}\times \bm{b} \cdot \bm{c}) = 0\,,
\end{equation}
to eliminate some of the quintuple products. The sign before individual terms in the last expression corresponds to the sign of the cyclic permutation of the four vectors.

The first class of $\omega$'s involves $\lp$ and $\lm$ in the mixed product, $\lm - \lp$ in the scalar product. Both products are even under $b \leftrightarrow \bar b$:
\begin{align}
    \omega_1 &\sim \left[(\lm \times \lp) \cdot \h\right] \left[(\lm - \lp)\cdot \h\right],\\
    \omega_2 &\sim \left[(\lm \times \lp) \cdot \h\right] \left[(\lm - \lp)\cdot (\lm+\lp)\right],\\
    \omega_3 &\sim \left[(\lm \times \lp) \cdot \h\right] \left[(\lm - \lp)\cdot (\b+\bbar)\right],\\
    \omega_4 &\sim \left[(\lm \times \lp) \cdot (\b+\bbar)\right] \left[(\lm - \lp)\cdot \h\right],\\
    \omega_5 &\sim \left[(\lm \times \lp) \cdot (\b+\bbar)\right] \left[(\lm - \lp)\cdot (\lm+\lp)\right],\\
    \omega_6 &\sim \left[(\lm \times \lp) \cdot (\b+\bbar)\right] \left[(\lm - \lp)\cdot (\b+\bbar)\right].
\end{align}
The second class involves $\b \times \bbar$ and/or $\b - \bbar$ in both mixed and scalar products:
\begin{align}
      \omega_7 &\sim \left[(\b \times \bbar) \cdot \h \right] \left[(\b - \bbar)\cdot \h\right],\\
      \omega_8 &\sim \left[(\b \times \bbar) \cdot \h \right] \left[(\b - \bbar)\cdot (\lm+\lp)\right],\\
      \omega_9 &\sim \left[(\b \times \bbar) \cdot \h \right] \left[(\b - \bbar)\cdot (\b+\bbar)\right],\\
      \omega_{10} &\sim \left[(\b \times \bbar) \cdot (\lm+\lp) \right] \left[(\b - \bbar)\cdot \h\right],\\
      \omega_{11} &\sim \left[(\b \times \bbar) \cdot (\lm+\lp) \right] \left[(\b - \bbar)\cdot (\lm+\lp)\right],\\
      \omega_{12} &\sim \left[(\b \times \bbar) \cdot (\lm+\lp) \right] \left[(\b - \bbar)\cdot (\b+\bbar)\right],\\
  \omega_{13} &\sim \left[(\b \times \bbar) \cdot (\lm-\lp) \right] \left[(\b - \bbar)\cdot (\lm-\lp)\right],\\
{\omega_{14}} &\sim {\left[(\lm \times \lp) \cdot (\b - \bbar)\right] \left[(\b - \bbar)\cdot (\lm-\lp)\right]}\,.
\end{align}
The third class involves mixed product of $\h$, $\lm \pm \lp$, and $\b \pm \bbar$:
{
\begin{align}
  \omega_{15} &\sim \left[\h\times (\lm + \lp) \cdot (\b - \bbar) \right] \left[(\b - \bbar)\cdot \h\right],\\
  \omega_{16} &\sim \left[\h\times (\lm + \lp) \cdot  (\b - \bbar) \right] \left[(\b - \bbar)\cdot (\lm+\lp)\right],\\
  \omega_{17} &\sim \left[\h\times (\lm + \lp) \cdot  (\b - \bbar) \right] \left[(\b - \bbar)\cdot (\b+\bbar)\right],\\
                \omega_{18} &\sim \left[\h\times (\lm - \lp) \cdot (\b + \bbar) \right] \left[(\lm - \lp)\cdot \h\right],\\
  \omega_{19} &\sim \left[\h\times (\lm - \lp) \cdot (\b + \bbar) \right] \left[(\lm - \lp)\cdot (\lm+\lp)\right],\\
  \omega_{20} &\sim \left[\h\times (\lm - \lp) \cdot (\b + \bbar) \right] \left[(\lm - \lp)\cdot (\b + \bbar)\right],\\
  \omega_{21} &\sim \left[\h\times (\lm - \lp) \cdot (\b - \bbar) \right] \left[(\lm - \lp)\cdot (\b - \bbar)\right].
\end{align}
}

There are further possibilities with a mixed product $\h \times (\lm + \lp)\cdot(\b +\bbar)$ multiplied by one of 8 $C$-even, $b\leftrightarrow \bar b$ even scalar products $\{ \h \cdot \h, \h \cdot (\lm + \lp), \h \cdot (\b + \bbar), (\lm + \lp)\cdot (\b + \bbar), (\lm \pm \lp)^2, (\b \pm \bbar)^2\}$. Since the mixed product itself already has the desired symmetry properties, those 8 quintuple products do not bring additional new information, with respect to a mixed (triple) product that is our final observable:
{
\begin{equation}
  \label{eq:omegaEx14Now22}
  \omega_{22} \sim \h \times (\lm + \lp) \cdot (\b +\bbar).
\end{equation}
}

Note that there are nonlinear relations between $\omega$'s, such as $\omega_1 \omega_6 = \omega_3 \omega_4$, that we do not exploit to further reduce the set. Namely, ratios of $\omega$'s can contain singularities in the available phase-space and as such would
be difficult to reconstruct by a neural network optimizer.

{All the $\omega$'s are normalized by the lengths of the vectors that enter as factors in the scalar products,
  \begin{equation}
    \label{eq:norm}
    \omega_i = \frac{[\bm{V}_1\times \bm{V}_2 \cdot \bm{V}_3]\,[\bm{V}_4\cdot \bm{V}_5] }{|\bm{V}_1\times \bm{V}_2||\bm{V}_3|\,|\bm{V}_4| |\bm{V}_5|}\,,
  \end{equation}
  and the upper bound $|\omega_i| \leq 1$ is generally valid. For cases when
  $\omega_i$ has a vector $\bm{a}$ present both in the mixed
  and scalar products, e.g.
  $(\bm{V}_1\times \bm{V}_2 \cdot \bm{a})\, (\bm{V}_3\cdot\bm{a})$, and
  furthermore with $\bm{V}_1 \times \bm{V}_2 \cdot \bm{V}_3 = 0$, a stricter
  upper bound $|\omega_i| \leq 1/2$ applies (for $\omega_{1,6,7,11,13,14,15,16,20,21}$).}

%\section*{References}

\bibliographystyle{elsarticle-num}
\bibliography{NNtop-ref}{}
\end{document}